# Magnetic Field Control of the Quantum Chaotic Dynamics of Hydrogen Analogues in an Anisotropic Crystal Field

Weihang Zhou,[1] Zhanghai Chen,[1,2,*] Bo Zhang,[2] C. H. Yu,[2] Wei Lu,[2] and S. C. Shen[2,1,†]

[1]*Surface Physics Laboratory, Department of Physics, and Laboratory of Advanced Materials, Fudan University, Shanghai 200433, P. R. China*
[2]*National Laboratory for Infrared Physics, Shanghai Institute of Technical Physics,
Chinese Academy of Sciences, Shanghai 200083, P. R. China*
(Dated: March 9, 2010)

We report magnetic field control of the quantum chaotic dynamics of hydrogen analogues in an anisotropic solid state environment. The chaoticity of the system dynamics was quantified by means of energy level statistics. We analyzed the magnetic field dependence of the statistical distribution of the impurity energy levels and found a smooth transition between the Poisson limit and the Wigner limit, *i.e.* transition between regular Poisson and fully chaotic Wigner dynamics. Effect of the crystal field anisotropy on the quantum chaotic dynamics, which manifests itself in characteristic transitions between regularity and chaos for different field orientations, was demonstrated.

The control and anti-control of chaos have been the subject of growing interdisciplinary interest in the past decades [1, 2]. The possibility of controlling chaos is not only interesting from the point of view of fundamental physics, but also of immense practical importance. A number of techniques have been developed to control chaos since the seminal work of Ott, Grebogi, and Yorke [3], many of which have been verified experimentally and found applications in a wide variety of fields, *e.g.* in electronic circuits, non-linear optics, biological systems *etc* [1, 2, 4, 5]. On the other hand, the presence of chaos in quantum mechanics (sometimes referred to as quantum chaos) has been demonstrated and can be found in quantum systems [6, 7]. A large amount of work has been stimulated since its introduction in 1970s. However, understanding the underlying physics of such quantum chaotic systems and finding ways of controlling them remain a major scientific challenge in this field.

Highly excited Rydberg atoms in strong magnetic fields are particularly suitable for studying the quantum manifestation of underlying chaotic dynamics, among which hydrogenic atoms received the most attention due to their physical simplicity and experimental feasibility [8]. However, our understanding of chaotic atoms is still far from being complete, not to mention chaotic atom analogues. The quantum manifestation of underlying chaotic dynamics and its control mechanism for atom analogues in solid state environment remain mysterious, despite their fundamental importance.

It is now generally known that atom analogues can be well produced in a solid state environment [6]. Perhaps, the physically most appealing conservative system exhibiting underlying chaotic dynamics is the analogue of the hydrogen atom in a semiconductor crystal. Studies have shown that hydrogen-like shallow impurities in semiconductor resemble the energy level structures of the hydrogen atom but with a much smaller energy scale due to the small electron effective mass and large dielectric constants [6, 9, 10]. The impurity electron feels the external field much more sensitively than in real atoms and thus efficiencies for the control of chaos can be greatly enhanced. Another significant difference from real atoms in vacuum is that series of material parameters can be manipulated, *e.g.* the electron effective mass tensor in a crystal of Si is anisotropic. Such analogues in solid state environment are ideal systems for studying the control of quantum chaotic dynamics, paving the way for the future application of chaos in semiconductor devices. And it is experimentally feasible to prepare a Rydberg-atom-like impurity in semiconductors and investigate its chaotic nature, as well as the corresponding chaos control mechanism, by means of spectroscopy techniques. However, little work has been reported.

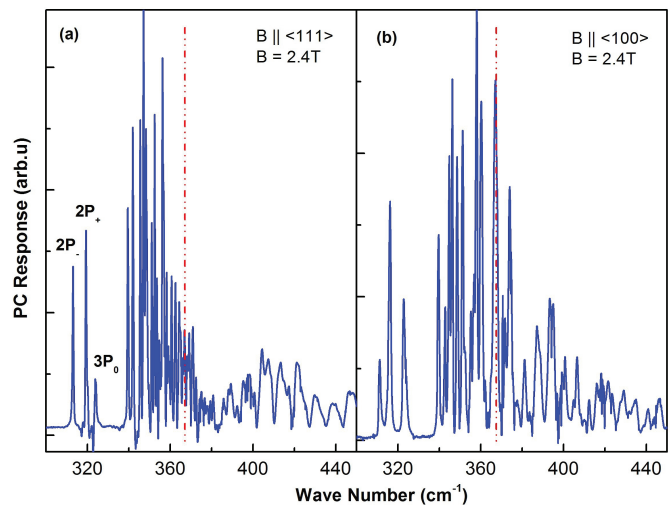

FIG. 1: Typical photo-thermal ionization spectroscopy of phosphorus impurities in Si taken under magnetic field of 2.4 T. (a) $\boldsymbol{B} \parallel k \parallel$ <111>. (b) $\boldsymbol{B} \parallel k \parallel$ <100>. The red dash-dot-dot line denotes the field-free ionization threshold ( $\sim 367$ cm$^{-1}$).

In this letter, we report experimental studies on the magnetic field control of the quantum chaotic dynamics of phosphorus shallow donor impurities in ultra-pure single crystalline silicon, by means of energy level statistics. To the best of our knowledge, this is the first investigation on quantum chaos control, as well as spectral fluctuations of the hydrogen analogue in an anisotropic solid state environment. The phosphorus concentration in our samples is estimated to be $10^{11}$ cm$^{-3}$, which is low enough that all donors in the sample can be regarded as isolated centers even for highly excited Rydberg states. Considering the anisotropic nature of the effective mass tensor of Si, we prepared two sets of samples for different experimental configurations: one for $B \parallel k \parallel <111>$ and the other for $B \parallel k \parallel <100>$, with $k$ being the wave vector of the incident infrared radiation and $B$ the applied magnetic field. The experimental technique we employed here is the so-called photo-thermal ionization (PTI) spectroscopy, which has been proven to be particularly suitable for detecting ultra-low concentration impurities in semiconductor due to its high sensitivity and high resolution [11]. All measurements were taken at 17 K to get the best experimental sensitivity and signal-to-noise ratio.

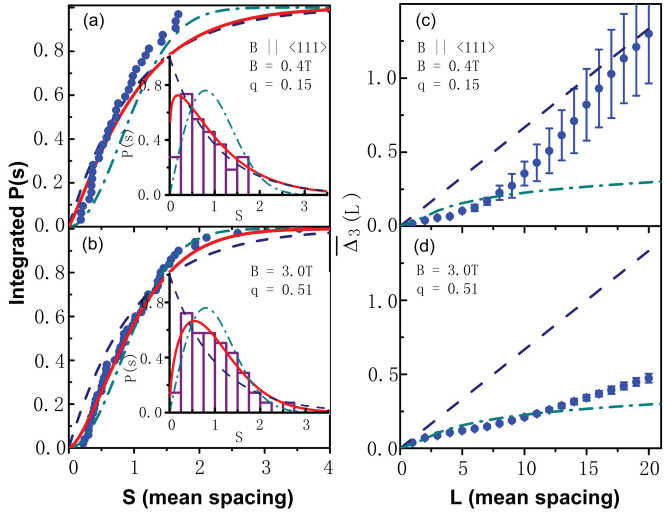

FIG. 2: (a) The integrated nearest-neighbor-spacing (NNS) distribution taken at magnetic field of 0.4 T for $B \parallel k \parallel <111>$ ; (b) the integrated NNS distribution taken at magnetic field of 3.0 T for $B \parallel k \parallel <111>$. The insets show the corresponding NNS distributions. Together shown in the figure are predictions from RMT for regular Poisson (dashed curve) and chaotic Wigner (dash-dotted) distributions. The red solid curve denotes the best-fit Brody distribution. (c) spectral rigidity and predictions from RMT for 0.4 T, $B \parallel k \parallel <111>$; (d) spectral rigidity and predictions from RMT for 3.0 T, $B \parallel k \parallel <111>$.

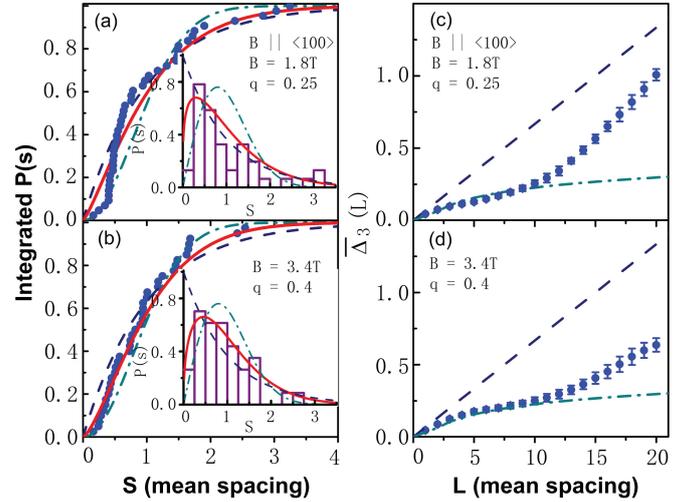

FIG. 3: (a) The integrated NNS distribution taken at magnetic field of 1.8 T for $B \parallel k \parallel <100>$ ; (b) the integrated NNS distribution taken at magnetic field of 3.4 T for $B \parallel k \parallel <100>$. The insets show the corresponding NNS distributions. Together shown in the figure are predictions from RMT for regular Poisson (dashed curve) and chaotic Wigner (dash-dotted) distributions. The red solid curve denotes the best-fit Brody distribution. (c) spectral rigidity and predictions from RMT for 1.8 T, $B \parallel k \parallel <100>$; (d) spectral rigidity and predictions from RMT for 3.4 T, $B \parallel k \parallel <100>$.

It is now widely recognized that the regular or chaotic nature of the classical dynamics of a bound Hamiltonian system manifests itself in the statistical fluctuations of the energy levels: the level distribution of an integrable system exhibits a clustering nature while level repulsion effect can be found in a quantum system with underlying chaotic dynamics [12–16]. There exists an established theory to describe the spectral fluctuation properties of a Hamiltonian system, namely the random matrix theory (RMT), which is introduced to nuclear physics in the 1960s by Wigner and developed mainly by Dyson and Mehta [12–14, 17]. RMT predicts that the nearest-neighbor-spacing distribution $P(s)$, which denotes the probability of finding two neighboring levels with a separation between $s$ and $s + ds$, is given by a Poisson distribution $P_{Poisson}(s) = \exp[-s]$ for an integrable system, while a time-reversible-invariant system with fully chaotic underlying classical dynamics is characterized by the Wigner Gaussian orthogonal ensemble (GOE) distribution $P_{Wigner}(s) = \pi s/2 \exp[-\pi s^2/4]$. For systems with mixed regular and chaotic dynamics, their nearest-neighbor-spacing distribution can be given by the heuristic Brody distribution $P_{Brody}(s, q) = (q + 1)\alpha s^q \exp[-\alpha s^{q+1}]$, $\alpha = [\Gamma\{(q+2)/(q+1)\}]^{1+q}$, where $q$ is the Brody chaoticity parameter that interpolates between the regular Poisson distribution ($q = 0$) and the chaotic Wigner distribution ($q = 1$) and can be roughly regarded as the percentage of the chaotic volume in its corresponding classical phase space [14].

First we investigate the spectral fluctuations in the case of $B \parallel k \parallel <111>$. Typical PTI spectrum of the phosphorus impurities in silicon taken at magnetic field of 2.4 T is shown in FIG. 1(a). Series of dense line structures are clearly visible, below and slightly above the zero-field ionization threshold ( $\sim 367$ cm$^{-1}$). These sharp peaks originate from the discrete shallow donor impurity levels, from which we carried out our level statistics analyses. Further above the zero-field ionization threshold, regular modulations of the spectra appear. They are the well-known quasi-Landau resonance patterns [18].

In a standard RMT statistical analysis, a sufficiently long and complete level sequence with identical quantum numbers is required, to ensure the statistical significance. However, spectra obtained experimentally are usually incomplete: instrumental resolution is limited and missed levels are often unavoidable. Moreover, they are probably not pure: levels with uncertain or incorrect quantum number assignments are often included. Relevant methods have been developed to account for such situations [12, 14, 19, 20]. In the case of $B \parallel k \parallel <111>$, the number of levels included in our analyses is about 50 for most field strength, with several lowest levels (those around 320 cm$^{-1}$) omitted to remove non-statistical fluctuations. Moreover, the sequence under analyses in our experiment is a mixed one: the sequence consists of six different m$^\pi$ (m: azimuthal quantum number; π: the z parity) series, determined by the optical selection rule. Thus, we focus our attention on the way the chaoticity parameter $q$ changes with the applied magnetic field, rather than its absolute value.

We now examine the nearest-neighbor-spacing distribution in the case of $B \parallel k \parallel <111>$. Considering the limited levels involved, we calculate the integrated nearest-neighbor-spacing distribution of $I(s) = \int_0^s P(s)ds$, to reduce noise. This can be done by counting the number of spacings smaller than $s$, and divide it by the total number of spacings included. Two representative integrated nearest-neighbor-spacing distributions taken at magnetic fields of 0.4 and 3.0T were shown in FIG. 2 (a) and (b). Promising agreement between the experimental data and their best-fit Brody distribution can be found. Together shown in the inset is the nearest-neighbor-spacing distribution itself. Agreements between the experimental data and the Brody fit are also promising.

The value of the chaoticity parameter $q$ is determined by fitting the heuristic Brody curve to the integrated nearest-neighbor-spacing distribution by means of a least-square-deviation procedure. The magnetic field dependence of the obtained $q$ was shown in FIG. 4. (a). The fitted parameter $q$ shows a rapid increase from nearly 0 at 0.2 T ($q = 0.01$, precisely) to a saturation value of about 0.45, implying that the underlying classical dynamics of the impurity electron undergoes a transition from regularity to chaos. This transition is indeed straightforward: at small fields, the magnetic field can be regarded as a perturbation and the system is near-integrable, leading to a near-Poisson level distribution; while at high fields, the magnetic field effect is no longer negligible, thus leading to a non-separable Hamiltonian and a near-Wigner level distribution.

While the short-range correlation between adjacent levels can be measured using the nearest-neighbor-spacing distribution, the long-range correlation between levels can be given by the spectral rigidity measurement $\Delta_3(L)$ [21]. For a fixed energy interval $[\alpha, \alpha + L]$, the Dyson-Mehta statistics $\Delta_3(L)$ is defined as the least-square deviation of the staircase function $N(E)$ from its best-fit straight line: $\Delta_3(L) = (1/L) Min_{A,B} \int_\alpha^{\alpha+L} [N(E)-AE-B]^2 dE$ (Here, $N(E)$ denotes the cumulative energy level count). For a given spectrum, the smaller $\Delta_3(L)$, the stronger is the spectral rigidity. The calculated rigidity for the two representative examples shown in FIG. 2 (a) and (b) has been given in FIG. 2 (c) and (d), respectively. As the external magnetic field is increased, the $\Delta_3(L)$ statistics exhibits a clear tendency moving towards the Wigner RMT prediction, showing again the transition from regularity to chaos. It should also be noted that the $\Delta_3(L)$ statistics are quite closed to the Wigner RMT prediction at small $L$, implying that correlation still preserves when two energy levels are not apart from each other far enough.

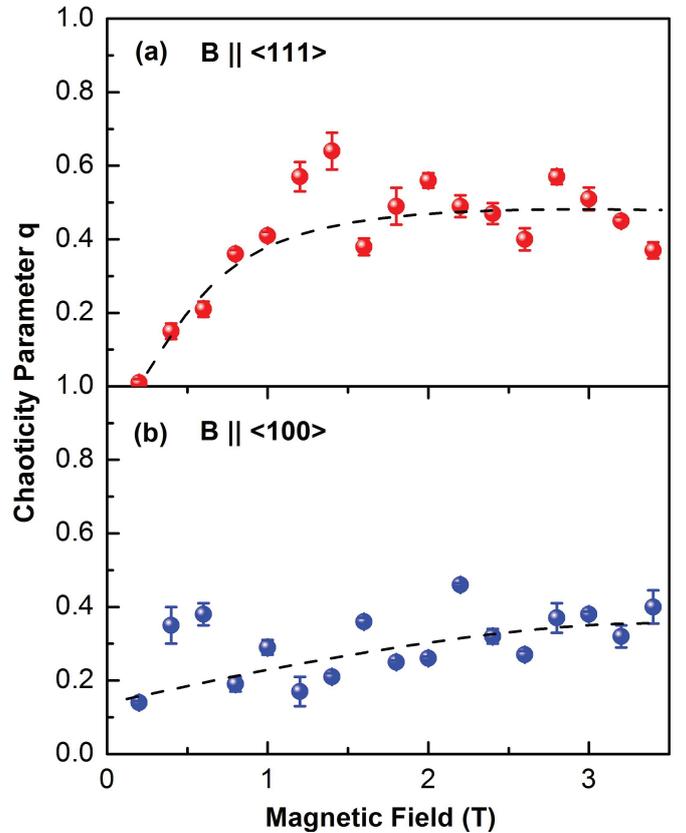

FIG. 4: Calculated chaoticity parameter $q$ under magnetic fields ranging from 0.2 to 3.4 T with a step of 0.2 T. (a) $B \parallel k \parallel <111>$; (b) $B \parallel k \parallel <100>$.

As has been stated previously, the electron effective mass tensor of silicon is anisotropic. To investigate how the anisotropic solid state environment affects the quantum chaotic dynamics and statistical level distributions of the impurity electrons, we now turn to the experimental configuration of $B \parallel k \parallel <100>$. Typical PTI spectrum is shown in FIG. 1(b). Similar to the case of $B \parallel k \parallel <111>$, dense peaks are clearly visible in the spectrum. The nearest-neighbor-spacing distributions, together with the corresponding spectral rigidity, have been calculated. The number of levels involved is about 50 for most fields and again several lowest levels (levels around 320 cm$^{-1}$) have been excluded. Representative examples taken at magnetic fields of 1.8 and 3.4T are shown in FIG. 3. Again, good agreements between the experimental data and the heuristic Brody fit were found. Consistency between the short-range (nearest-neighbor-spacing distribution) and long-range (Dyson-Mehta statistics) correlation measurements is promising.

The value of the chaoticity parameter $q$ has been calculated using the same procedure as in the case of $\boldsymbol{B} \parallel k \parallel <111>$, with its details plotted in FIG. 4 (b). Global tendency towards larger $q$, i.e. a transition from the Poisson limit to the Wigner limit, is found again. However, Difference between the two experimental configurations is significant. Compared with the chaoticity parameter $q$ in the case of $\boldsymbol{B} \parallel k \parallel <111>$ which shows clearly rapid increase and fast saturation, $q$ in the configuration of $\boldsymbol{B} \parallel k \parallel <100>$ exhibits a rather slow increase with a large fluctuation as the external field increases. At high fields near 3.0 T, $q$ reaches a saturation value of about 0.3. Such significant difference can be well interpreted as follows: The crystal field of Si is anisotropic. There are six equivalent ellipsoids in its constant-energy-surface, which produce one unique electron effective mass of $m^* \approx 0.28 m_e$ for $\boldsymbol{B} \parallel k \parallel <111>$ and two different effective masses of $m_1^* \approx 0.19 m_e$ and $m_2^* \approx 0.42 m_e$ for $\boldsymbol{B} \parallel k \parallel <100>$. Thus, the level sequence in the case of $\boldsymbol{B} \parallel k \parallel <100>$ is actually a superposition of two classes of electrons characterized by their different effective masses. Such a superposition obscures the tendency of the way $q$ changes with magnetic field, leading to the slow increase of the chaoticity parameter $q$ for $\boldsymbol{B} \parallel k \parallel <100>$. Moreover, numerical simulations have shown that the superposition of several statistically independent Wigner sequences would lead to a decrease of the chaoticity parameter $q$, i.e. a transition towards the Poisson limit [12, 14, 19, 20]. The more the sequences are mixed, the smaller is the parameter $q$. In the case of six mixed Wigner sequences, the level distribution of the combined sequence would be rather closed to the Poisson limit. However in our experiment, the saturation value of $q$ is much larger than zero (0.45 for $\boldsymbol{B} \parallel k \parallel <111>$ and 0.3 for $\boldsymbol{B} \parallel k \parallel <100>$). This may imply that the different $m^\pi$ sequences in our case are not completely statistically independent. It should also be pointed out that the quasi-Landau resonance observed above the zero-field ionization threshold has been widely accepted as signature of underlying quantum chaotic dynamics [6, 18, 22, 23]. The gradually strengthened quasi-Landau resonance pattern with the applied magnetic field is thus an evident reflection of the underlying classical dynamics moving towards chaos [18].

In summary, we report the first experimental study, by means of energy level statistics, on the magnetic field control of the quantum chaotic dynamics of hydrogen analogues in an anisotropic solid state environment. The short-range nearest-neighbor-spacing distribution, together with the long-range spectral rigidity, was analyzed for the phosphorus shallow impurity in silicon. The ultra-low donor concentration and ultra-high experimental resolution and sensitivity ensure the high confidence level of the resolved shallow impurity energy levels. Promising agreements between the experimental data and the heuristic Brody fit were found. The dependence of the statistical distribution of the impurity energy levels on the external magnetic fields have been carefully examined and a smooth transition between the Poisson limit and the Wigner limit, i.e. transition between regular and fully chaotic underlying classical dynamics, was demonstrated for both field orientations. Effect of the anisotropic nature of the crystal field manifests itself in characteristic transitions between regularity and chaos for different field orientations. Indeed, the transition between regular and quantum chaotic dynamics of the impurity electrons can be well controlled by an external magnetic field. Our work thus provides direct evidence for the quantum manifestation of the underlying chaotic dynamics and chaos control mechanisms for the impurity electrons in solid state environment, giving strong implication for the future application of chaos in semiconductor devices.

The work is financed by NSFC and 973 projects of China (No.2006CB921506). We thank Prof. C. M. Hu and Prof. Donglai Feng for their useful suggestions.

* Electronic address: zhanghai@fudan.edu.cn
† Electronic address: xcshen@fudan.edu.cn
[1] S. Boccaletti, C. Grebogi, Y. -C. Lai, et al., Phys. Rep. **329**, 103 (2000).
[2] S. Boccaletti, J. Kurths, G. Osipov, et al., Phys. Rep. **366**, 1 (2002).
[3] E. Ott, C. Grebogi, J. A. Yorke, Phys. Rev. Lett. **64**, 1196 (1990).
[4] E. Scholl, Nature Phys. **6**, 161 (2010).
[5] S. Steingrube, M. Timme, et al., Nature Phys. **6**, 224 (2010).
[6] M. C. Gutzwiller, *Chaos in Classical and Quantum Mechanics* (Springer-Verlag New York Inc, New York,1990).
[7] P. B. Wilkinson, T. M. Fromhold, et al., Nature **380**, 608 (1996).
[8] H. Friedrich, D. Wintgen, Phys. Rep. **183**, 37 (1989).
[9] R. A. Faulkner, Phys. Rev **184**, 713 (1969).
[10] S. C. Shen, J. B. Zhu, Y. M. Mu, P. L. Liu, Phys. Rev. B **49**, 5300 (1994).
[11] S. M. Kogan, T. M. Liftshits, Phys. Stat. Sol. (a) **39**, 11 (1977)
[12] H. A. Weidenmuller, G. E. Mitchell, Rev. Mod. Phys **81**, 539 (2009).
[13] T. Guhr, et al., Phys. Rep. **299**, 189 (1998).
[14] T. A. Brody, et al., Rev. Mod. Phys **53**, 385 (1981).
[15] D. Ullmo, et al., Phys. Rev. Lett. **90**, 176801 (2003).
[16] J. Main, et al., Phys. Rev. A **57**, 1149 (1998).
[17] M. L. Mehta, *Random Matrices and the Statistical Theory of Energy Levels* (Academic, New York, 1967).
[18] Z. H. Chen, W. H. Zhou, et al., Phys. Rev. Lett. **102**, 244103 (2009).
[19] O. Bohigas, M. P. Pato, Phys. Lett. B **595**, 171 (2004).
[20] O. Bohigas, M. P. Pato, Phys. Rev. E **74**, 036212 (2006).
[21] F. J. Dyson, M. L. Mehta, J. Math. Phys. **4**, 701 (1963).
[22] A. Holle, et al., Phys. Rev. Lett. **61**, 161 (1988).
[23] J. Main, et al., Phys. Rev. Lett. **57**, 2789 (1986).